\documentclass[epj]{svjour}
\usepackage[utf8]{inputenc}
\usepackage[T1]{fontenc}
\usepackage{amssymb}
\usepackage{amsmath}
\usepackage{amsbsy}
\usepackage{bm}
\usepackage{graphicx}
\usepackage{url}
\usepackage{hyperref}
\hypersetup{
    colorlinks=true,
    linkcolor=blue,
    filecolor=blue,      
    citecolor=blue,
    urlcolor=blue,
}
\newcommand{\I}{\mathrm{i}}
\newcommand{\E}{\mathrm{e}}

\begin{document}
\title{Edge states in a two-dimensional quantum walk with disorder}
\author{Alberto D.\ Verga\inst{}\thanks{E-mail: Alberto.Verga@univ-amu.fr}}
\institute{%
Aix Marseille Université, Centre de Physique Théorique, Campus de Luminy, 13288 Marseille, France}   

\date{\today}
\abstract{
We investigate the effect of spatial disorder on the edge states localized at the interface between two topologically different regions. Rotation disorder can localize the quantum walk if it is strong enough to change the topology, otherwise the edge state is protected. Nonlinear spatial disorder, dependent on the walker's state, attracts the walk to the interface even for very large coupling, preserving the ballistic transport characteristic of the clean regime.
}
\maketitle

%
\section{Introduction}

Thirty years ago Feynman devised quantum algorithms using unitary transformations of an initial quantum state \cite{Feynman-1986kx}. In 1993, Aharonov et al.  \cite{Aharonov-1993fk} introduced the notion of quantum walk through a coin and a shift operator in analogy with the classical random walk, and found that information propagates at a ballistic rate instead of a diffusive one. Also in the nineties, Meyer \cite{Meyer-1996sf} defined quantum cellular automata in one particle sector, and demonstrated that its continuous limit leads to a Dirac equation in two dimensions. The genealogy of quantum walks shows their rich physical content, ranging for quantum information to condensed matter \cite{Kempe-2003fk,Venegas-Andraca-2012zr,Kitagawa-2012fk,Bianconi-2015uo}.

Quantum walks are especially interesting because they provide an original point of view of quantum systems based on the properties of quantum states instead of the more usual approach based on the energy levels and eigenstates of a Hamiltonian. This point of view shed a new light on the dynamics of quantum systems and the mechanisms by which they explore the available Hilbert space. At variance to the standard definition of a quantum system by its Hamiltonian, a quantum walk is defined by an evolution unitary operator. In particular, the relationship between the quantum state and information allows to relate concepts from quantum information theory to material systems through the introduction of a quantum walk effective Hamiltonian. We may also invert the reasoning and ask whether it is possible to get some insight on the behavior of quantum walks using concepts from condensed matter. 

Some classes of quantum walks possess remarkable topological properties whose origin can be traced back to the structure of their evolution operator which links the particle's spin (coin operator) with its momentum (shift operator), akin to the spin-orbit coupling in solid state, leading to nontrivial Berry phases, edge states and non vanishing Chern numbers \cite{Kitagawa-2010jk,Asboth-2012qy,Asboth-2013yq}. Experiments \cite{Schreiber-2011qf,Svozilik-2012qy,Xue-2015ys} show that in the presence of disorder, the quantum walk losses its coherent interference pattern and, depending on the nature of the noise, can localize or transit to a classical diffusion regime. Noise also affects the topological phases and the localization-delocalization properties of one dimensional \cite{Rakovszky-2015kx} and two dimensional quantum walks \cite{Edge-2015yq}. In the absence of disorder, a one dimensional quantum walk can still localize at the interface between two distinct topological regions \cite{Kitagawa-2012xy}. These phenomena are also characteristic of topological phases in condensed matter, like quantum Hall \cite{Prange-1990xy} or spin Hall insulators \cite{Hasan-2010fk} where edge states appear. However, it is worth noting that such bound states are protected against disorder. As a consequence, the transport properties (quantization of the conductance, for instance) are preserved as long as the topology is not changed.

Our aim is to investigate the effect of spatial noise on the edge states localized at the interface between two regions differing in their topology. 

%

\section{Quantum walk topology and effective Hamiltonian}

\begin{figure}
  \centering
  \includegraphics[width=0.9\linewidth]{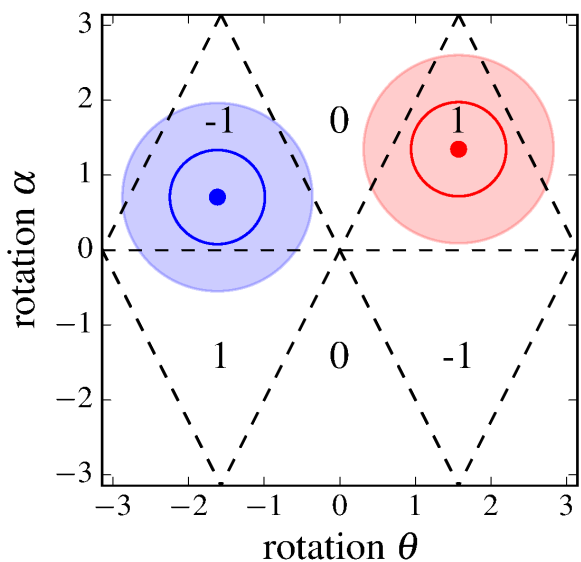}
  \caption{Phase diagram $(\theta,\alpha)$, the triangular sectors corrspond to different Chern numbers $C$ (see Ref.~\protect\cite{Kitagawa-2010jk}). Rotation angles for $x<0$ (red point, $C=1$) and $x>0$ (blue point, $C=-1$). The circles represent the distribution of random angles for $J=0.1$ (inner circle) and $J=0.2$ (outer circle). In the second case the noise may change the topology of the quantum walk.
  \label{f:c-phases-I}}
\end{figure}

We consider a discrete time quantum walk in a square lattice, with nonzero Chern number \cite{Kitagawa-2010jk}. The walker's Hilbert space is the set of spinors \(|\psi(t)\rangle\) depending on time \(t\). They are given by the Kronecker product \(\otimes\) of the walker's position and spin state \(|\bm x\rangle \otimes |s\rangle\), where \(\bm x = (x,y) \in \mathbb{Z}^2\) is a lattice node (the lattice constant \(a=1\) is the length unit), and \(s = \uparrow, \downarrow\) takes ``up'' and ``down'' values. The state evolution is determined by a ``coin'' \(R\) operator acting in spin space, and a ``shift'' \(T\) operator. At a given time step, the motion direction depends on the spin orientation. At the end of the walk, a measure of the position density probability distribution \(P(x,y,t)\), is performed.

\begin{figure}
  \centering
  \includegraphics[width=0.9\linewidth]{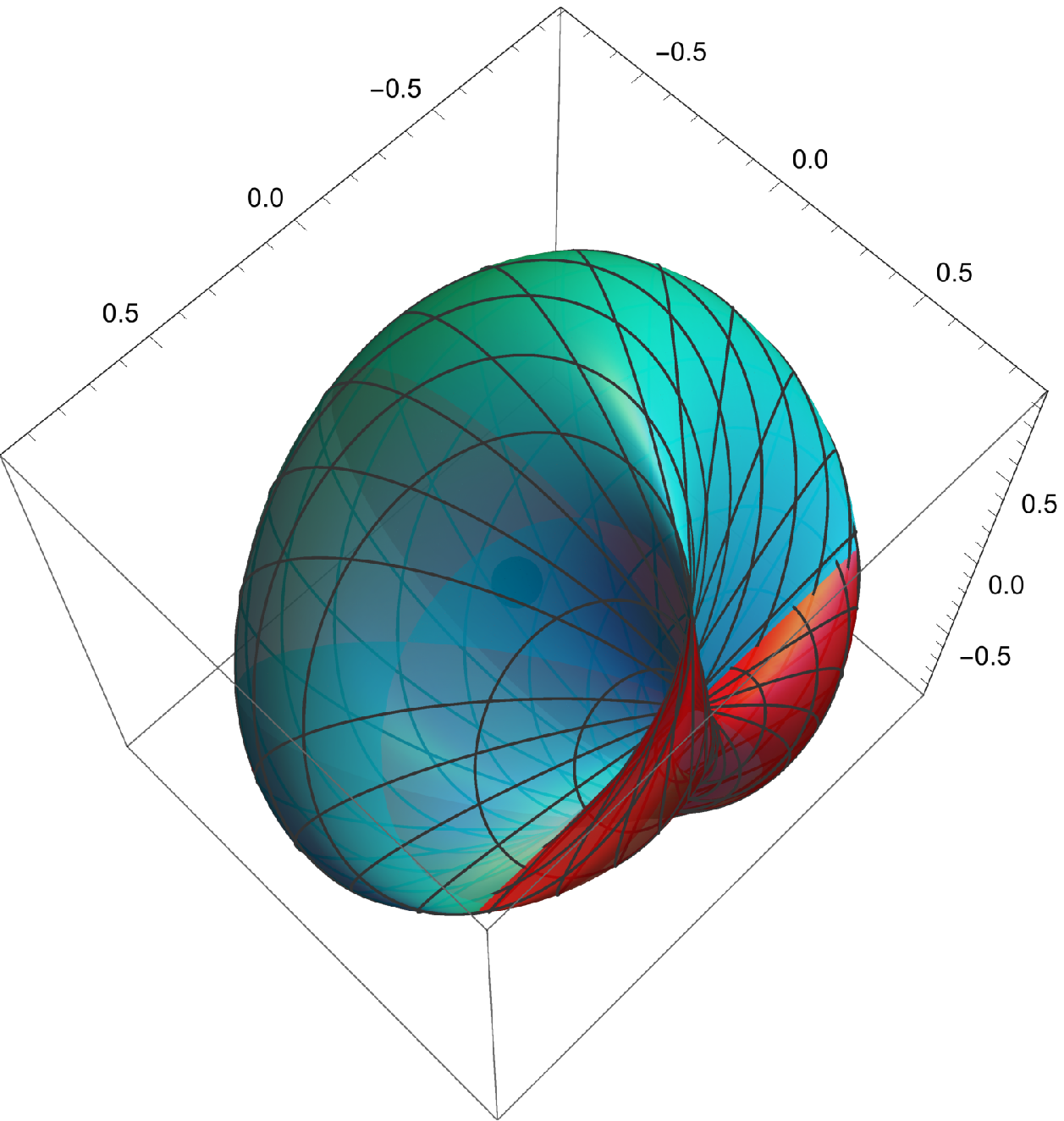}\\
  \includegraphics[width=0.9\linewidth]{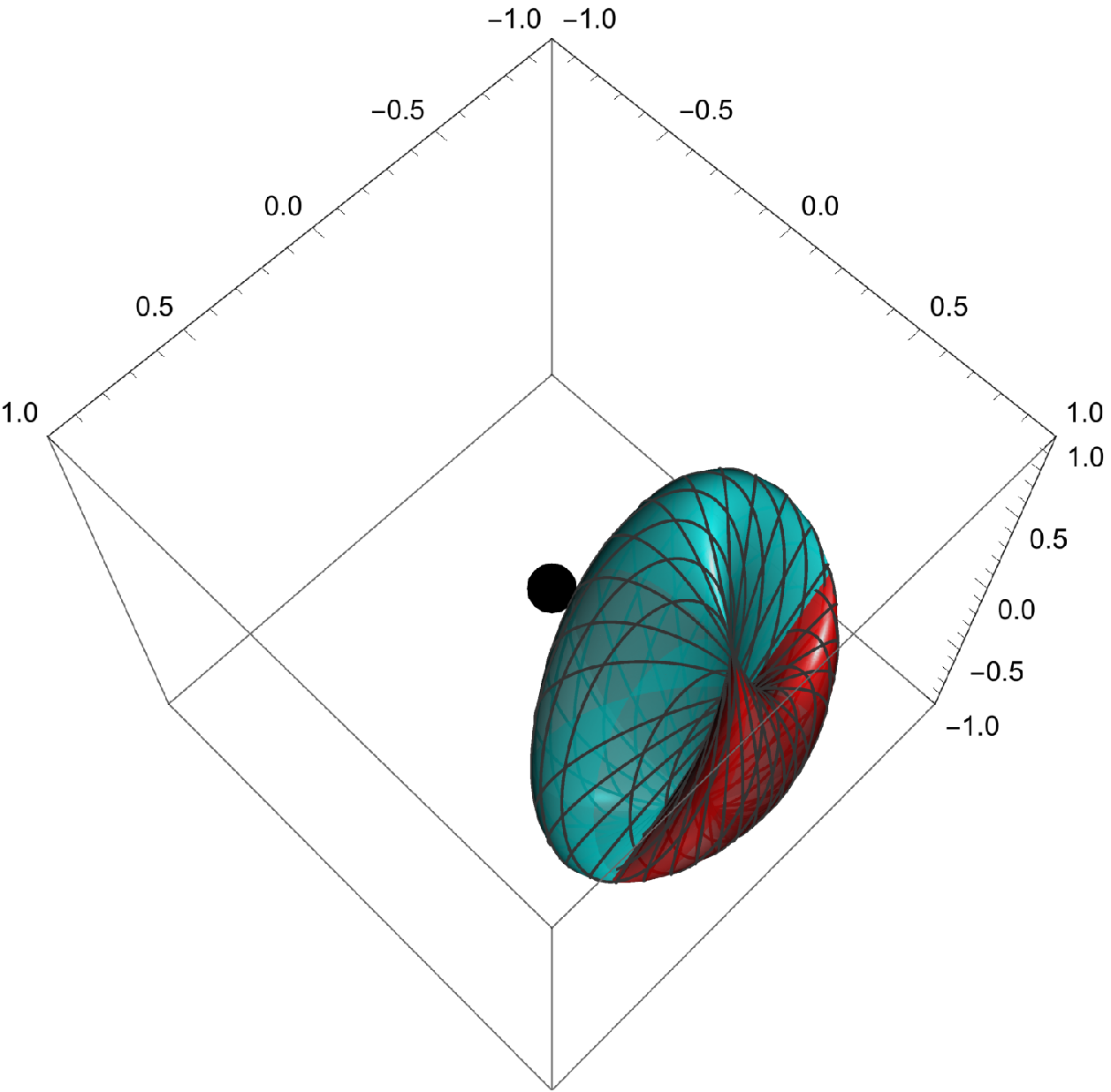}
  \caption{The topology of the effective Hamiltonian $H=E(\bm k) \bm d(\bm k)\cdot \bm \sigma$ change when the vector $\bm d(\bm k)$ vanishes. In the top panel $\theta=\pi/2$, $\alpha=3\pi/7$, the $\bm d$ vector surrounds the origin (black point), which corresponds to a nontrivial topology; in the bottom panel $\theta=\sqrt{2}/2$, $\alpha=(1+\sqrt{5})/2$, the topology is trivial. The two colors correspond to the inner and outer surfaces.
  \label{f:c0}}
\end{figure}

The initial state,
\begin{equation}
  \label{e:is}
|\psi(0)\rangle = |0\rangle \otimes \frac{1}{\sqrt{2}} \big(
| \uparrow \rangle + \I | \downarrow \rangle \big)\,,
\end{equation}
is a superposition of spin up \(\psi_\uparrow\) and down \(\psi_\uparrow\) amplitudes:
$$\psi_\uparrow = 1/\sqrt{2}\,,\quad \psi_\downarrow = \I/\sqrt{2}\,,$$
given equal probabilities of the spin orientation in the $z$ direction. The coin operator rotates the spin around the $y$ axis by an angle \(\theta\),
\begin{equation}
  \label{e:Rt}
  R(\theta) = 1_{\bm x} \otimes \E^{-\I \theta \sigma_y/2} = 
    1_{\bm x} \otimes \begin{pmatrix}
    \cos \theta/2 & -\sin \theta/2 \\
    \sin \theta/2 & \cos \theta/2
  \end{pmatrix} \,,
\end{equation}
where \(1_{\bm x}\) is the unit matrix in position space. The shift operator moves the particle to a neighboring node according to its spin projection:
\begin{multline}
  \label{e:Txy}
  T(p) = \sum_{\bm x} \big(
    |\bm x + \bm e_p, \uparrow\rangle \langle \bm x, \uparrow| + \\
    |\bm x - \bm e_p, \downarrow\rangle \langle \bm x, \downarrow| 
  \big) \,, \quad p = x,y
\end{multline}
where \(\bm e_p\) is a unit vector in the $p$ direction. One time step is executed by the unitary operator,
\begin{equation}
  \label{e:U6}
  U(\theta, \alpha) = T(x) R(\theta) T(y) R(\alpha) T(y)T(x)R(\theta)\,,
\end{equation}
where the product \(T(y)T(x)=T(x)T(y)\) shift the position by \(\pm 1\) on both directions \((x,y)\rightarrow(x \pm 1, y \pm 1)\).

\begin{figure}
  \centering
  \includegraphics[width=0.9\linewidth]{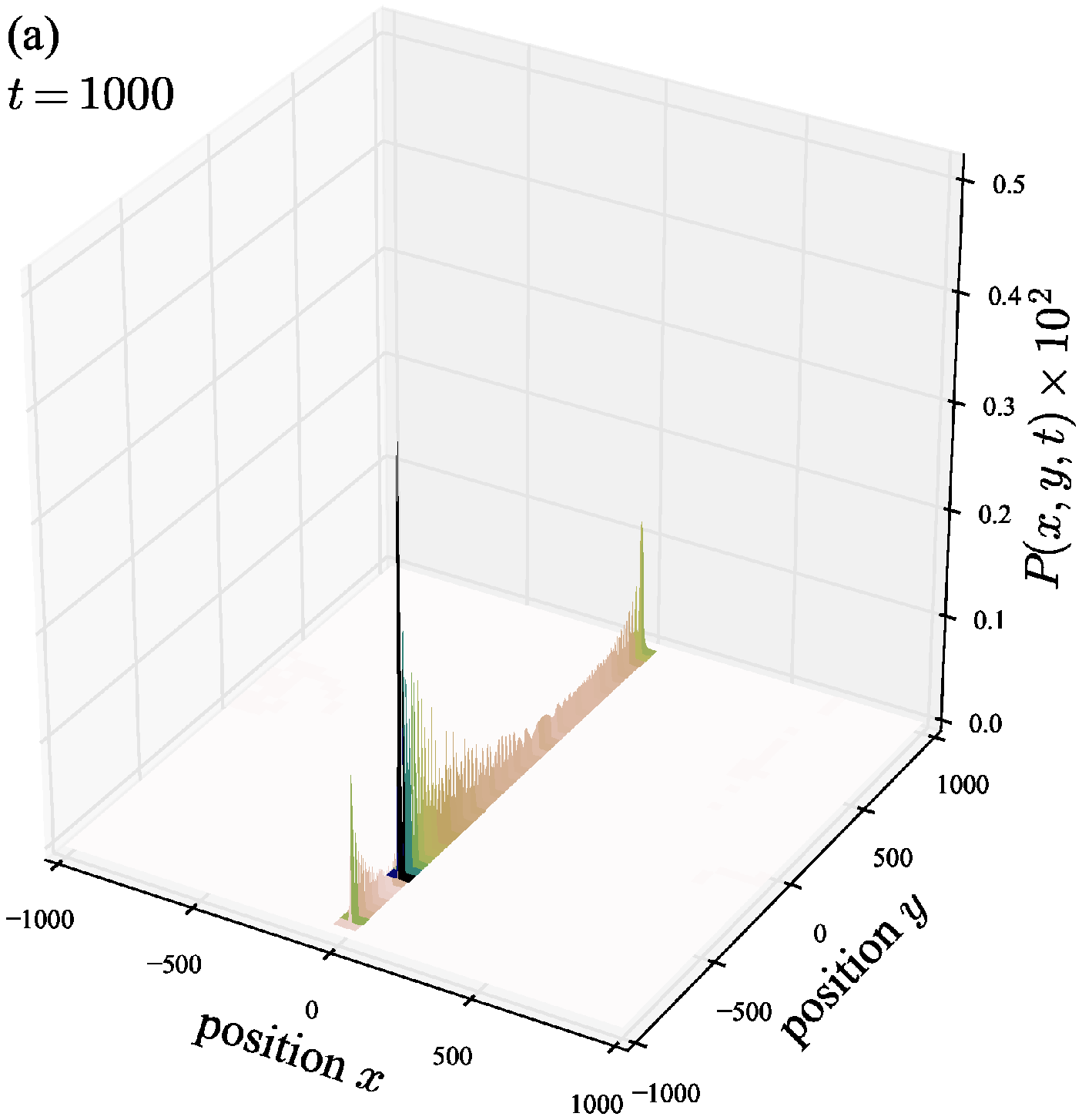}\\
  \includegraphics[width=0.9\linewidth]{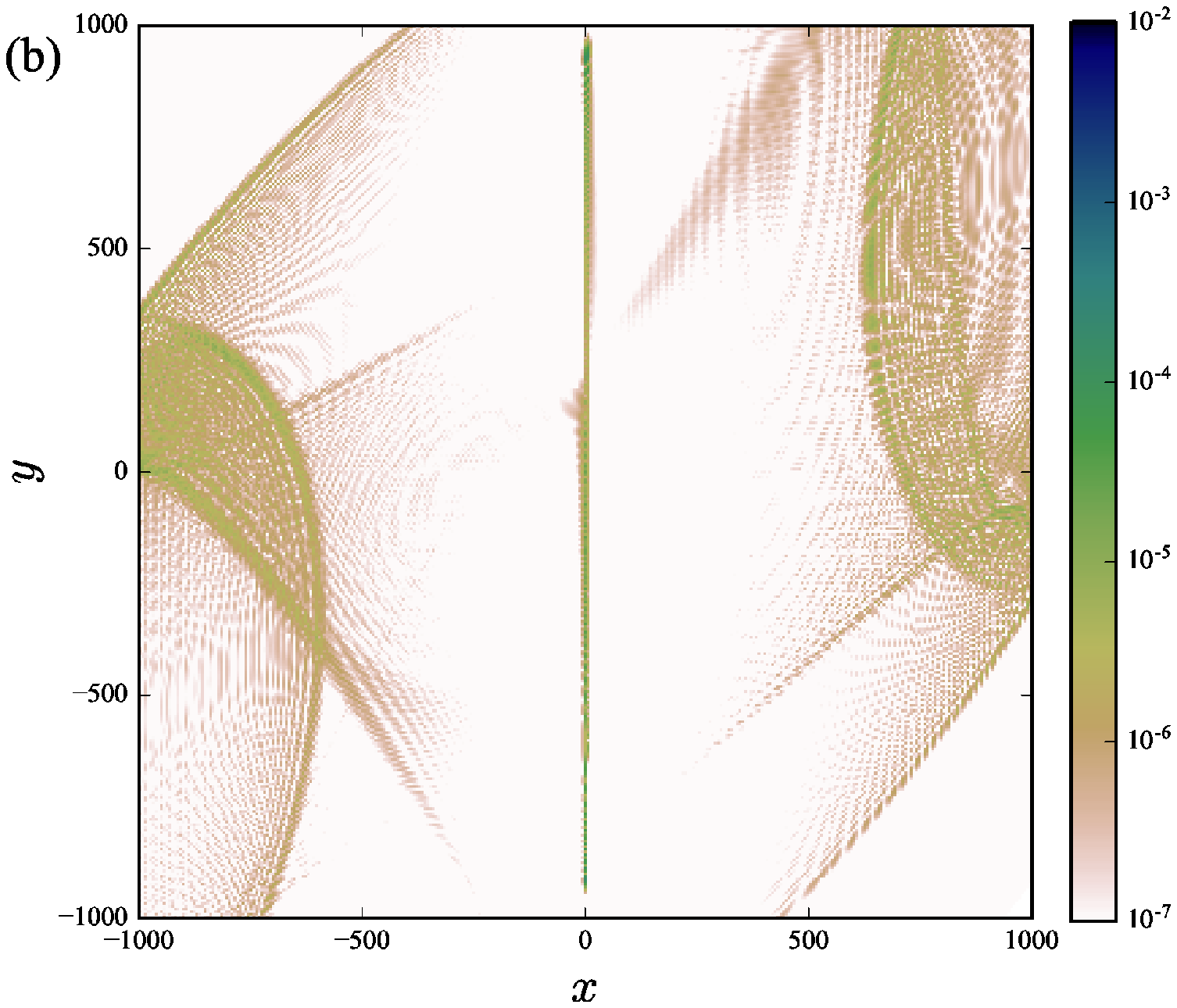}
  \caption{Position probability of a quantum walk with an interface between two regions having distinct topology: $P=|\psi_\downarrow|^2+|\psi_\uparrow|^2$. The top panel (a) shows that the walker distribution at time \(t=1000\), is essentially concentrated along the interface. The bottom panel (b) shows the same data in a logarithmic scale to reveal the structure of the small amplitudes (the wave folding due to the periodic boundary is not visible in a linear scale).
  \label{f:00}}
\end{figure}

According to the values of the rotation angles the family of effective Hamiltonians,
\[
  H(\theta, \alpha) = \I \log U(\theta, \alpha)  \,,
\]
is characterized by a Chern number taken the values \(C = \{-1,0,1\}\). The effective Hamiltonian is not uniquely defined, its eigenvalues are determined modulo \(2\pi\) (quasi-energies). The topological properties of the walk related to the symmetries of the effective Hamiltonian \cite{Kitagawa-2010jk}, depend on the values of the two parameters \((\theta,\alpha)\), as schematically represented in Fig.~\ref{f:c-phases-I}. It is worth noting that the topological classification of the quantum walks is richer than the one inferred from the sole effective Hamiltonian symmetries: taking into account the properties of the (one period) evolution operator implies the existence of a pair of topological invariants \cite{Obuse-2015ly}. As a consequence, topological protected edge states can appear at the interface of two trivial effective Hamiltonian phases \cite{Asboth-2012qy,Edge-2015yq}.

\begin{figure}[t]
  \centering
  \includegraphics[width=1.0\linewidth]{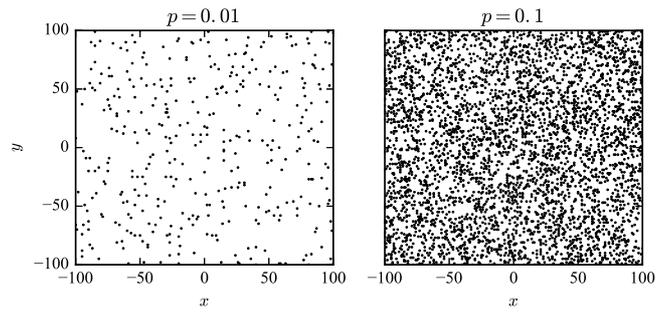}
  \caption{Random distribution of impurities for $p=0.01$ (left) and $p=0.1$ (right) over a square lattice of size $201\times201$.
  \label{f:c-impurities}}
\end{figure}

The explicit form of $H$ is easily obtained using the momentum representation \(\bm k = (k_x,k_y)\) of the unitary operator $U$, in the base of the Pauli matrices $\bm \sigma$:
\begin{equation}
  \label{e:uk}
  U(\bm k) = T(k_x)R(\theta)T(k_y)R(\alpha)T(k_x+k_y)R(\theta),
\end{equation}
where
\[
  R(\theta) = \cos\theta\, \sigma_0 - \I \sin\theta\, \sigma_y, \;
  T(k) = \cos k\, \sigma_0 + \I \sin k\, \sigma_z\,
\]
from which one readily obtains,
\begin{equation}
  \label{e:uks}
  U(\bm k) = d_0 \sigma_0 - \I\bm d\cdot \bm \sigma,\; \bm d = (d_x,d_y,d_z),
\end{equation}
or equivalently,
\begin{equation}
  \label{e:Heff}
  H(\bm k) = E(\bm k) \bm n(\bm k) \cdot \bm \sigma\,,
\end{equation}
where the energy spectrum is given by,
\begin{multline}
  \label{e:Ek}
  \cos E(\bm k) =
  \cos \frac{\alpha }{2} \big[
  \cos \theta  \cos k_x \cos(k_x+2 k_y)- \\
  \sin k_x \sin(k_x+2 k_y)\big]-
  \sin\frac{\alpha }{2} \sin\theta \cos^2 k_x\,,
\end{multline}
and the unit vector \(\bm n(\bm k) = \bm d(\bm k)/ d(\bm k)\), with
\begin{align}
  d_x &= \sin k_x 
  \big[
    \cos\frac{\alpha}{2} \sin\theta  \cos(k_x+2 k_y) - \notag
  \\
  & \phantom{{}=1}
    2 \sin\frac{\alpha}{2} \sin^2\frac{\theta}{2} \cos k_x
  \big] \,, \notag \\
  d_y &= \sin\frac{\alpha}{2} 
  \big(
     \cos\theta \cos^2k_x + \sin^2k_x \big) + \notag \\
     & \phantom{{}=1}
    \cos\frac{\alpha}{2} \sin\theta \cos k_x \cos(k_x+2 k_y) \,, \notag \\
  d_z &= \sin\frac{\alpha}{2} \sin\theta \sin k_x \cos k_x - \notag \\
    & \phantom{{}=1} \cos\frac{\alpha}{2} \big[
     \cos\theta \sin k_x \cos(k_x+2 k_y) + \notag \\
   & \phantom{{}=1}  \cos k_x \sin(k_x+2 k_y)\big] \,.
  \label{e:dk}
\end{align}
A quantum walk with effective Hamiltonian \eqref{e:Heff} possesses a nontrivial topology if the \(\bm n\) vector (or equivalently \(\bm d\)), surrounds the origin when the \(\bm k\) vector scans the Brillouin zone \([-\pi,\pi]^2\). In Fig.~\ref{f:c0} we represented two cases differing in their Chern number. We observe in the parametric plot of the vector \(\bm d\) over the Brillouin zone that depending on the choice of the pair \((\theta,\alpha)\), the origin is surrounded, signaling a nontrivial topology, or it is not surrounded.

The random walk thus defined possesses particle-hole symmetry but not time-reversal symmetry. Indeed, from the explicit form \eqref{e:Heff}, we verify that under conjugation \((\I \rightarrow -\I, \bm k \rightarrow -\bm k)\) the effective Hamiltonian changes sign: the one step operator \eqref{e:U6} is real implying that the walk has particle-hole symmetry. The walk defined by \eqref{e:U6} with particle-hole symmetry and broken time reversal symmetry, is thus reminiscent to class D in the usual classification \cite{Schnyder-2008yg}. 

\begin{figure*}[t]
  \centering
  \includegraphics[width=0.5\linewidth]{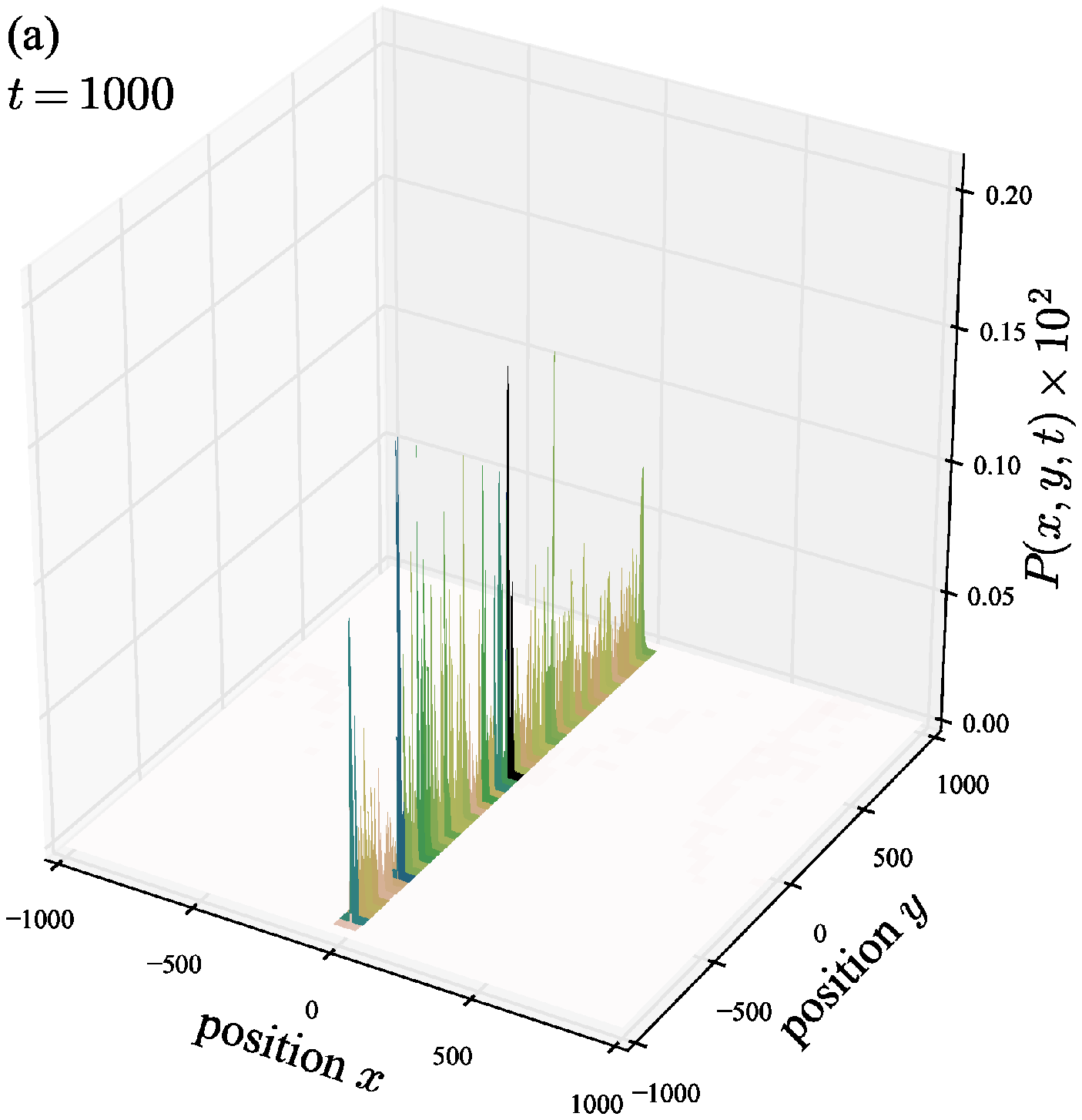}%
  \includegraphics[width=0.5\linewidth]{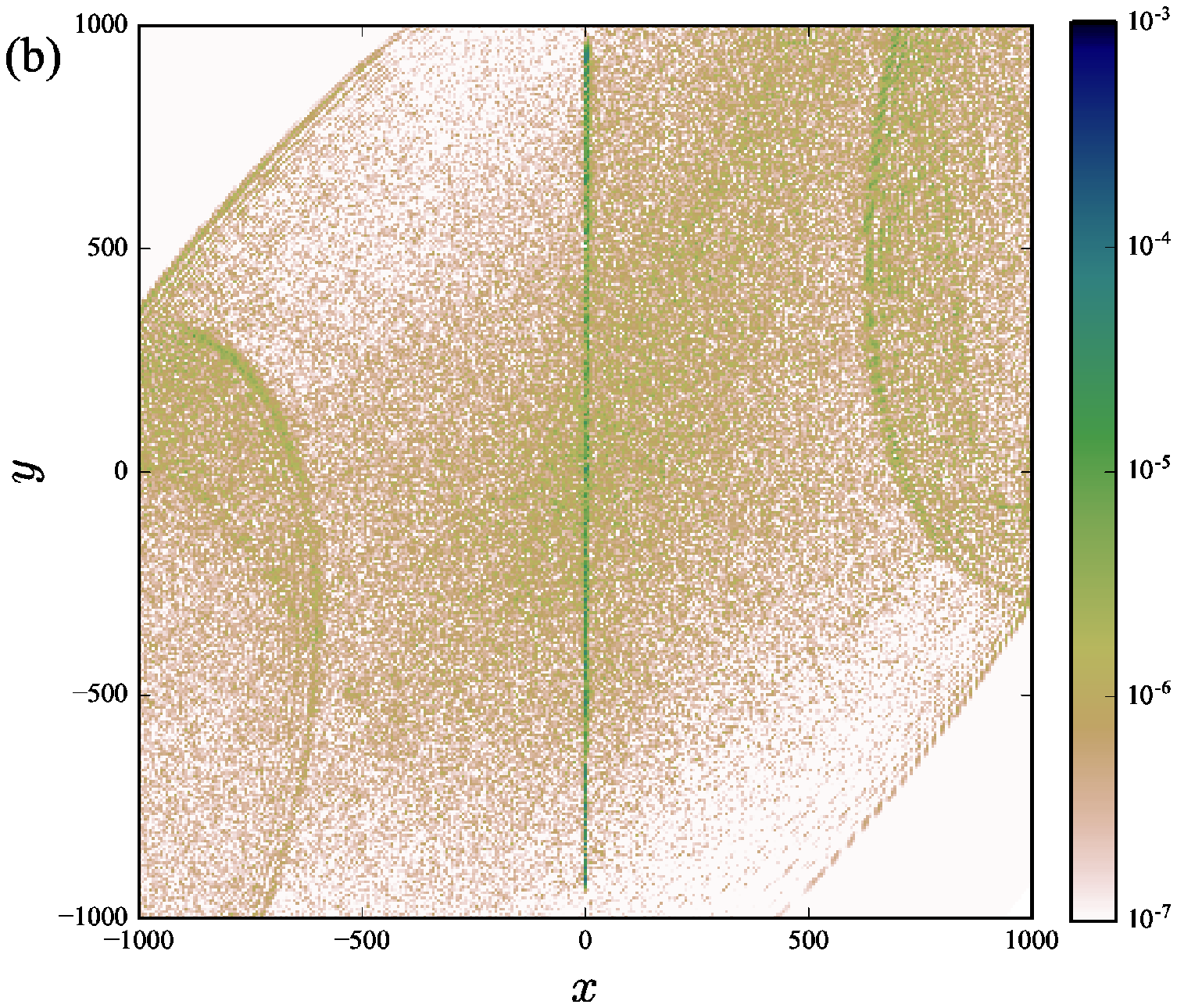}//
  \includegraphics[width=0.5\linewidth]{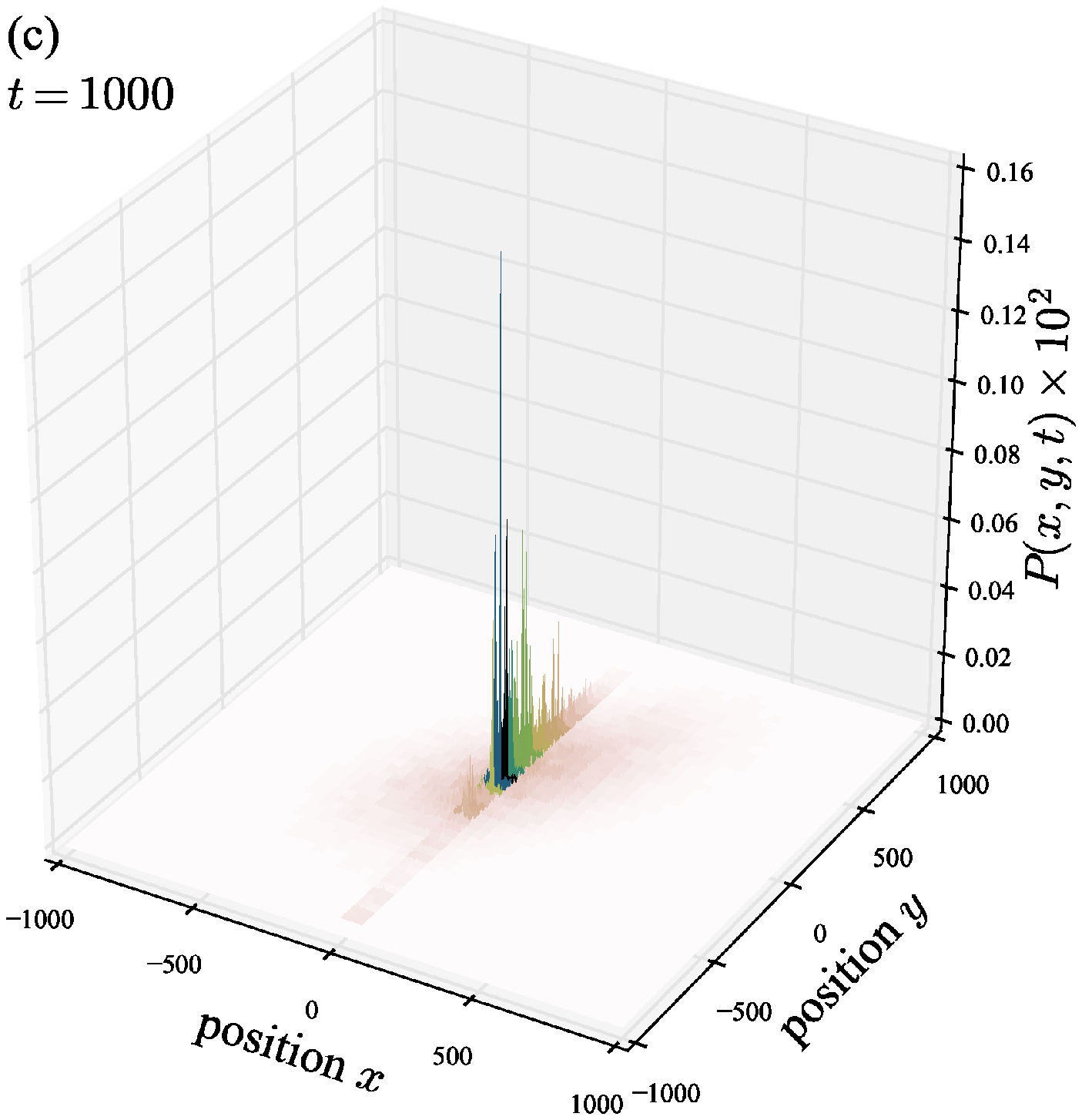}%
  \includegraphics[width=0.5\linewidth]{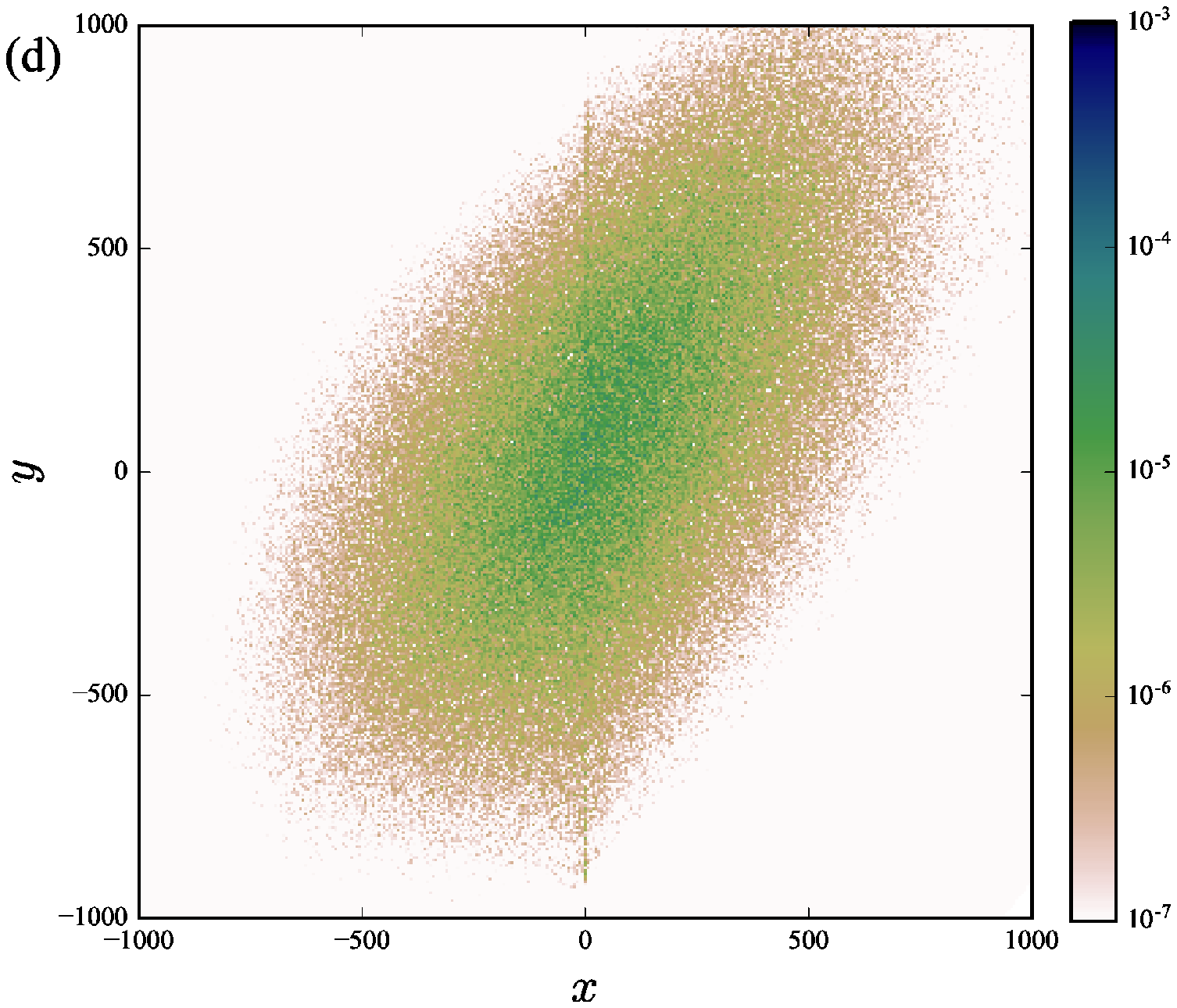}
  \caption{Position probability distribution at $t=1000$ for $p=0.03$ and $J=0.05$ (a-b) and $J=0.2$ (c-d). The coupling $J=0.2$ is strong enough to change the topology (c.f.\ \protect\ref{f:c-phases-I}); as a consequence the edge state is no longer protected and the walk localizes. Comparison with Fig.~\protect\ref{f:00} shows that in the low noise case $J=0.05$, the edge state is protected, but it is wider than in the clean case, and moreover, diffusion destroys the walk ballistic spreading outside the interface.
  \label{f:chern_12a}}
\end{figure*}

To investigate the consequences of the nontrivial topology, in particular the edge states, we perform a series of numerical computations for different disorder types. We use Eq.~\eqref{e:U6} to calculate the walker state evolution for one time step, in a square lattice with periodic boundary conditions. We split the lattice into two regions separated by an interface at \(x=0\), differing in their Chern number: for \(x\le 0\) we take \( (\theta,\alpha) = (\pi/2, 3\pi/7) )\), and for \(x>0\), \((\theta,\alpha)= (-(1+\sqrt{5})/2, \sqrt{2}/2)\), with \(C = 1,-1\), respectively (Fig.~\ref{f:c-phases-I}). In the one dimensional case the presence of an edge state at the origin leads to a localization of the quantum walk \cite{Kitagawa-2010jk,Kitagawa-2012xy}. In two dimensions one may expect propagation along the edge state localized at the interface. This is precisely what we observe, as demonstrated in Fig.~\ref{f:00} where a realization of the quantum walk \eqref{e:U6} is shown. The propagation along the interface is ballistic, as in the case of a one dimensional quantum walk. This phenomenon is observed for initial states having an overlap with the edge bound state, otherwise the quantum walk do not localize and explore the whole lattice with ballistic speed.

%
\section{Spatial disorder}

In order to investigate the effect of disorder we consider a set of randomly distributed sites \(I\), we call impurities, where the coin operator change. The spatial concentration of impurities is given by the probability \(p\) for a site to be occupied by an impurity. In Fig.~\ref{f:c-impurities} we show two of such distributions with $p=1\%$ and $p=10\%$. In such sites the simplest modification of the coin rule \eqref{e:Rt} is to make the angles \(\theta(\bm x) \) position dependent, 
$$
  \theta \rightarrow \theta(\bm x) = \theta + J \delta\theta(\bm x)\,, \quad
  \delta\theta \sim \mathcal{U}(0,2\pi)
$$
with \(\bm x \in I\) the set of impurity sites, and \(\mathcal{U}\) the uniform probability distribution in the given interval (i.e.\ rotation disorder). The parameter $J$ measures the strength of the disorder. For \(J\) small enough to let the system's topology unchanged, one may expect the edge state protected; in the opposite case one expects localization (or diffusive spreading in same special cases \cite{Edge-2015yq}) of the quantum walk (Fig.~\ref{f:c-phases-I}).

It is important to emphasize that the introduction of spatial disorder do not change the unitary evolution of the walker, nor the symmetries of the coin operator; however it modifies in a nontrivial manner the physics of the system by breaking the translational invariance, only preserved in a statistical sense, and by introducing new correlations between different sites that should change the walk interference patterns and hence its spreading properties. The effects of spatial noise in quantum walks are well studied, especially in one dimension; they can localize the walker or change its spreading rate \cite{Brun-2003nr,Ahlbrecht-2011kx,Schreiber-2011qf}. A more subtle behavior appears in topologically nontrivial walks, where localized zero energy modes may coexist with delocalized nonzero energy modes \cite{Obuse-2011fj}. We investigate in the case of two dimensions effects alike to this one.

\begin{figure}
  \centering
  \includegraphics[width=0.9\linewidth]{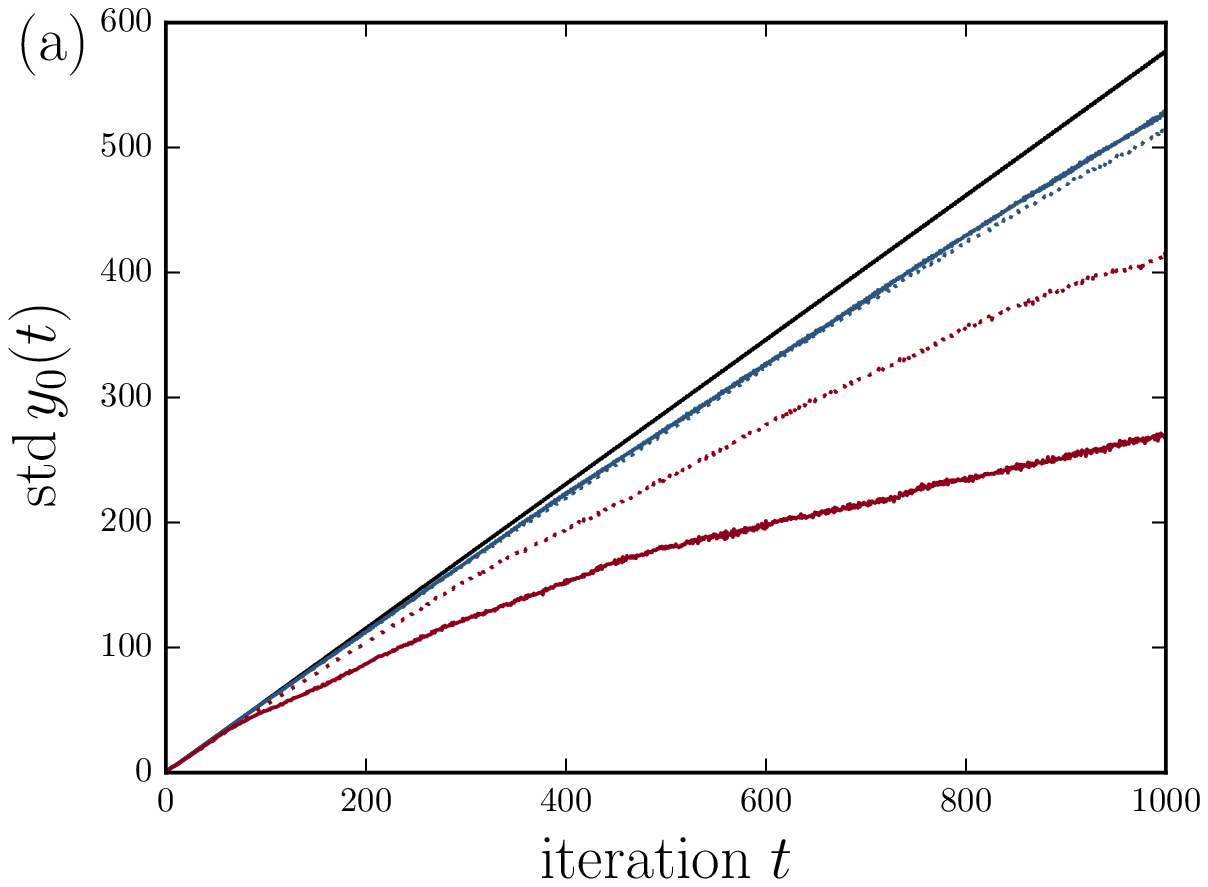}\\
  \includegraphics[width=0.9\linewidth]{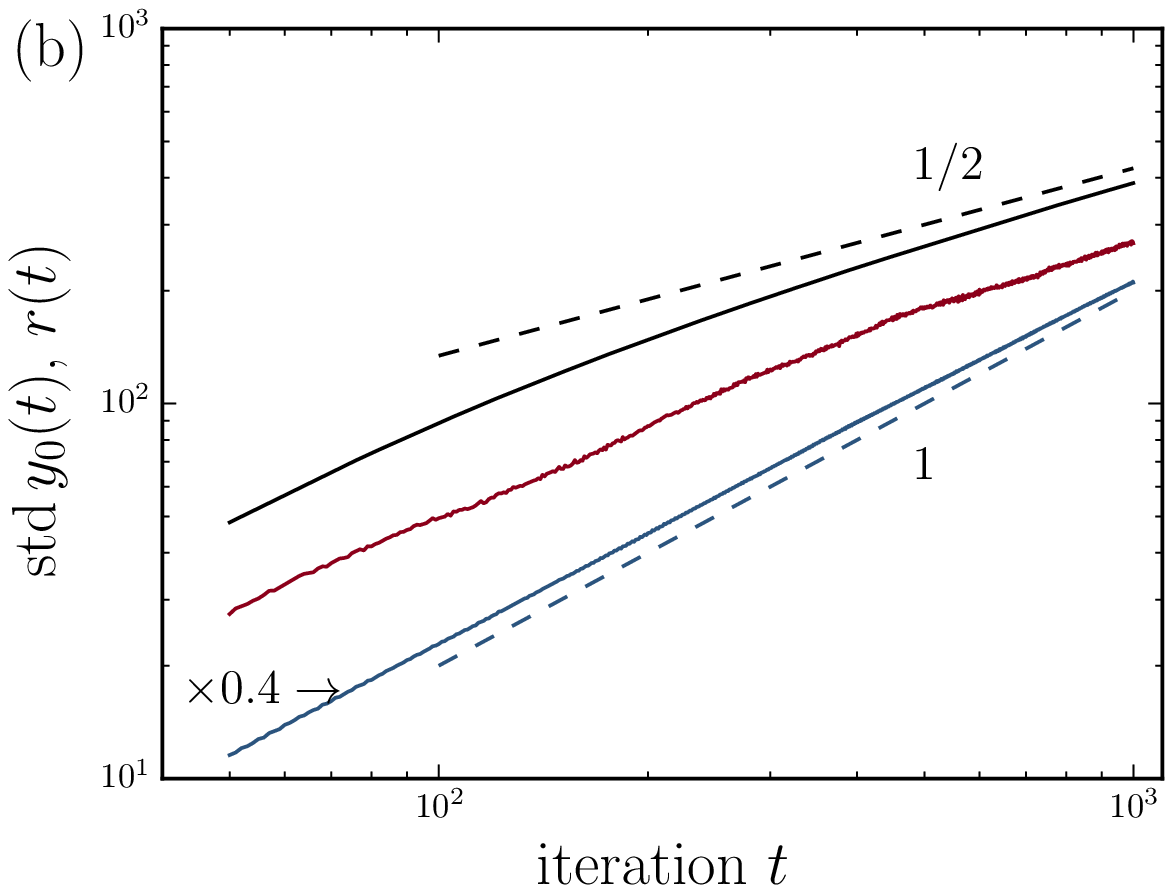}\\
  \includegraphics[width=0.9\linewidth]{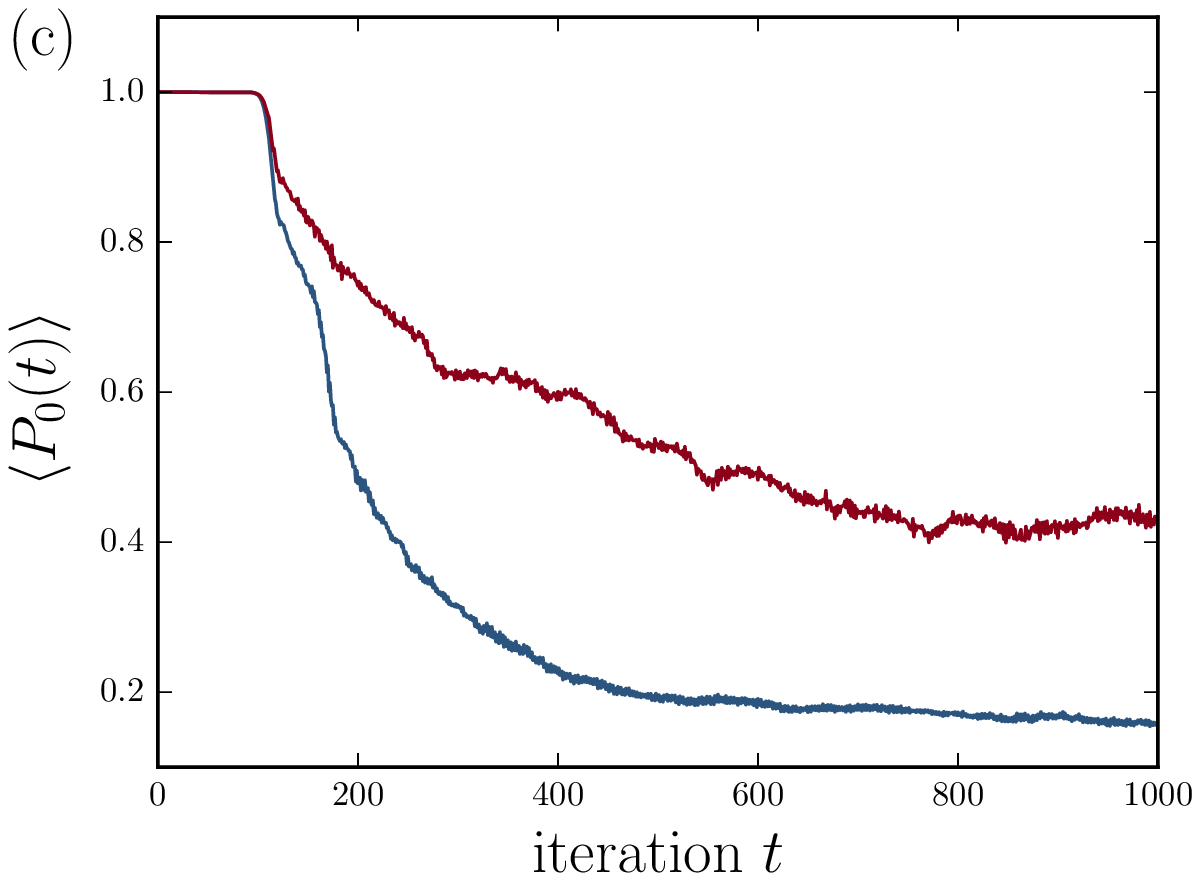}
  \caption{Quantum walk in the presence of rotation disorder. Width of the density distribution along the interface (a) as a function of time, for $J=0.05$ (blue lines) and $J=0.2$ (red lines), and $p=0.03$ (solid lines) and $p=0.01$ (dotted lines); the upper black line corresponds to the clean case. (b) Radial (black line) and at the interface  spreading in logarithmic scale; $J=0.05$ (blue, scaled by a factor $0.4$ to fit inside the frame), $J=0.2$ (red). The dashed lines correspond to the fitting exponents $1$ and $1/2$. For weak disorder the propagation is ballistic (exponent $1$), and approaches diffusion for strong disorder (exponent $1/2$). (c) Probability to be on the interface in a neighborhood of the origin, $x=0$ and $y\in (-100,100)$, $J=0.05$ (lower blue line), and $J=0.2$ (upper red line).
  \label{f:chern_p0}}
\end{figure}

We show a numerical computation of the random rotation quantum walk in Fig.~\ref{f:chern_12a}, to compare the weak and strong noise cases with the clean system of Fig.~\ref{f:00}. The presence of the interface introduces an anisotropy; actually, the evolution of the initial condition leads to an asymptotically inhomogeneous distribution of the position probability of the walker with different spreading properties on the interface and in the bulk. We find that the known general picture is confirmed, expectedly a localization transition is observed at a finite value of the disorder strength, as occurs in the one dimensional split-step quantum walk \cite{Rakovszky-2015kx}. The point is that in two dimensions, even if the interference pattern of the walk is lost away form the interface and the bulk distribution is localized or spread slowly, the ballistic transport along the edge is not affected for weak enough disorder. 

To quantify the effect of noise on the walker spreading, we measure the width of the probability density:
\begin{equation}
  \label{e:prob}
  P(\bm x,t) = |\psi_\uparrow(\bm x,t)|^2 + |\psi_\downarrow(\bm x,t)|^2
\end{equation}
using the definition 
\begin{equation}
  \label{e:width}
  w(t) = \sum_{\bm x} |\bm x|^2 P(\bm x,t) - 
  \left( \sum_{\bm x} \bm x P(\bm x,t) \right)^2
\end{equation}
which gives \(\mathrm{std}\,r(t) = \sqrt{w(t)}\), and an analogous definition for \(\mathrm{std}\,y_0(t)\) for the width on the interface \(x=0\):
\begin{equation}
  \label{e:stdy}
  \mathrm{std}\,y_0(t) = \left[
    \frac{\sum_y y^2 P(0,y,t)}{\sum_y P(0,y,t)}
  \right]^{1/2}\,. 
\end{equation}
(Note the time dependent normalization and the absence of mean term in order to catch running away distributions.)

Figure~\ref{f:chern_p0} shows the width for different values of the disorder strength $J$. Moreover we plot the probability to stay in a neighborhood of the origin \(P(x=0, y \in [-100,100],t)\) for weak and strong disorder. The measure of the isotropic width shows that the presence of disorder destroys the ballistic regime. However, the edge state transport remains ballistic for weak enough disorder. The width of the position distribution at the interface (\(x=0\)), increases linearly with time only for weak disorder. Concomitantly, the probability to find the walker near the origin tends to zero, in contrast to the persistence observed for stronger disorder. The deviation of $\mathrm{std}r(t)$ to a straight line together with the linear $\mathrm{std}y_0(t)$ is a manifestation of the underlying anisotropy of the walk in the weak noise regime, which tends to localize in the bulk and to propagate at the interface. 

\begin{figure}
  \centering
  \includegraphics[height=17em]{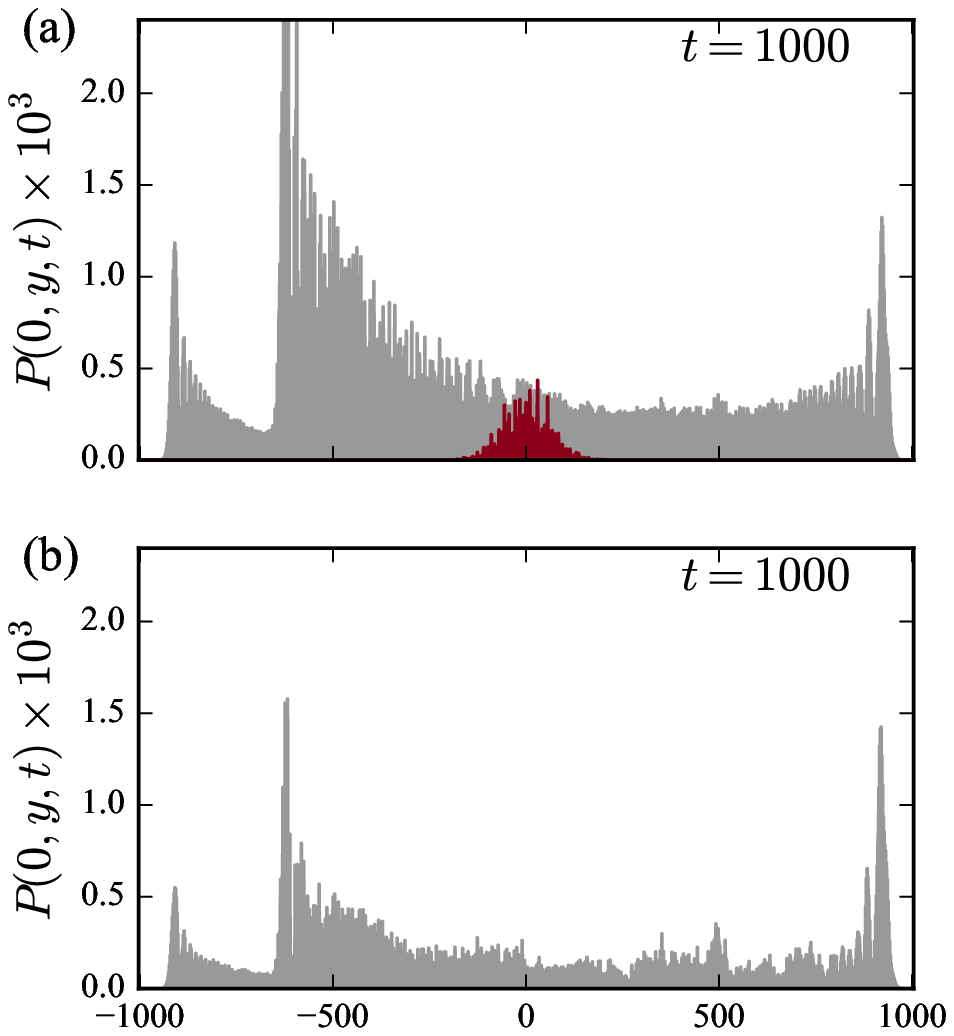}%
  \includegraphics[height=17em]{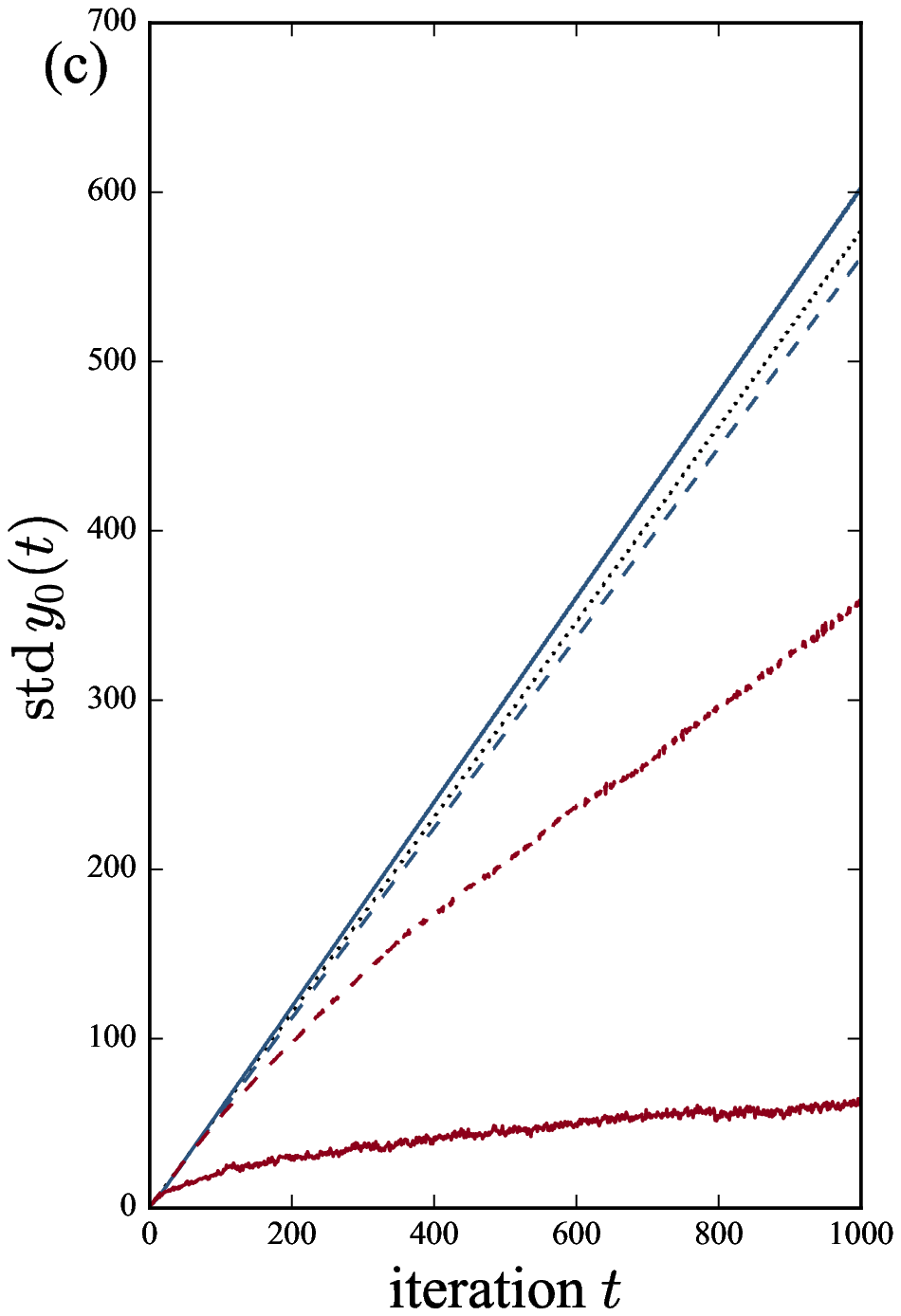}
  \caption{Nonlinear and phase disorder quantum walk. Probability distribution of the nonlinear walk, (a) for $J=1$ and (b) $J=10$,  ($p=0.1$, $t=1000$). In (a) superposed to the nonlinear walk (in gray), we show the random phase walk using the same parameters, localized near the origin (in red). (c) Width in the $y$ direction for $x=0$, at the interface, showing ballistic propagation for $J=0$ (dotted line), $J=1$ (blue dashed line), and $J=10$ (blue solid line), in the nonlinear case; one can compare with the random phase disorder (red dashed line) $J=0.1$ and (red solid line) $J=1$, showing diffusive spreading. Note the similar distribution in (a) with the one of the clean state (Fig.~\protect\ref{f:00}).
  \label{f:chern_0809}}
\end{figure}

\section{Nonlinear spatial disorder}

\begin{figure*}
  \centering
  \includegraphics[width=1.0\linewidth]{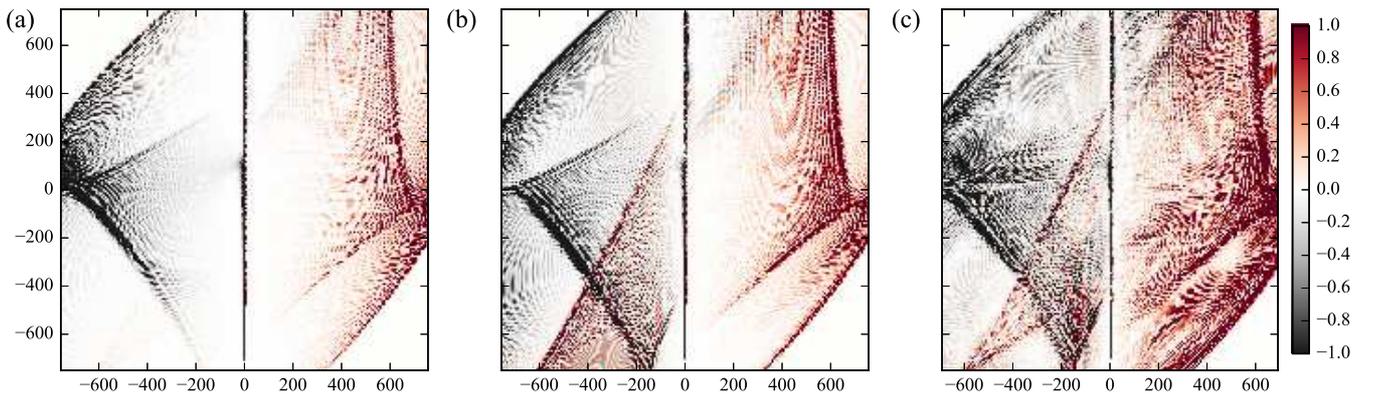}
  \caption{Zoom on the small scales of the spin density distribution $s\times10^6$,  $s=|\psi_\uparrow|-|\psi_\downarrow|^2$, for $J=0$ (a), $J=10$ (b), and $J=100$ (c) ($p=0.1$, $t=750$). In the nonlinear case (b-c), spin up and down waves are present on both sides of the interface.
  \label{f:chern_0809s}}
\end{figure*}

It is interesting to investigate the effect of disorder under a more complex setting, in particular generalizing the walk to take into account nonlinear (many-body) interactions \cite{Navarrete-Benlloch-2007fk,Molfetta-2015uq,Lee-2015uq,Gerasimenko-2016fk}. Pursuing our analogy with matter systems, we may think the impurities as being fixed spins whose interaction with the itinerant spin of the walker is of the exchange type. This ``magnetic disorder'' would introduce into the coin operator a phase factor \(\phi\)
\[
  S_J(\phi) = \E^{-\I J \phi(\bm x,t) \sigma_z}\, ,
\]
where \(\bm x \in I\) and $J$ the coupling constant, which leads to the redefinition
\begin{equation}
  \label{e:phi}
R(\theta) \rightarrow R_J(\theta,\phi) = S_J(\phi)R(\theta)\,.
\end{equation}
The simplest choice is to take $\phi$ as a random angle, uniformly distributed on the circle \cite{Edge-2015yq} (i.e.\ phase disorder). However, if we assume that the orientation of the fixed spin is determined by the orientation of the itinerant one (its $z$ spin component), \(\phi\) will be related to the walker's state, \(\phi = \pi s(\bm x, t)\):
\begin{equation}
  \label{e:s}
  s(\bm x,t) = |\psi_\uparrow(\bm x,t)|^2 - |\psi_\downarrow(\bm x,t)|^2\,,\;
  \bm x \in I\,.
\end{equation}
leading to a nonlinear coupling. The coin operator \(R_J(\theta,\phi)\) replaces $R$ in \eqref{e:U6}; it is now a composition of a rotation around the $y$ axis (with a uniform angle in each region) followed by a rotation around the $z$ axis at the occupied sites of the lattice. The spatial distribution of the defects remains random and we refer to the choice \eqref{e:phi,e:s} as ``nonlinear disorder''. However, the values of \(\phi\) at different sites are correlated by \eqref{e:s} with the walker's state, making the nonlinear disorder essentially different to the rotation or magnetic disorder types.

One important consequence of the introduction of the coin operator \eqref{e:phi} with respect to the random rotation one of the previous section, is that the form $R_J(\theta,\phi)$ breaks the so-called particle-hole symmetry. The unitary operator of the walker is no longer real: the rotation spin axis has now a $z$ component. As a matter of fact, the coin operator can be put in a form of a rotation around an axis depending on \(\phi\) of angle also depending on \(\phi\). This is true even if the angle $\phi$ is taken randomly instead of being related to the walker's state, although the implications on the existence of edge states in both cases, nonlinear disorder and random phases, can in principle strongly differ. The breaking of the particle-hole symmetry change the topological properties of the quantum walk putting it in the same class as the integer quantum Hall system (see Ref.~\cite{Kitagawa-2010jk}, Appendix B).  

To highlight the specific properties of the nonlinear walk, we chose the coupling parameter and the concentration in a range about $J\sim 1\,$--$\,100$, and $p\sim 0.1$, respectively. These large values of $J$ are justified because the spin density is of the order of $p$ and decreases at a given site at least as $1/t$ because of the probability spreading over the lattice (this is the scaling corresponding to the ballistic transport over the interface, supposing no spreading in the transverse direction). Significant results are obtained for values with $J>1/p$; the probability density on the interface scales as \(P \sim 1/t\) in the ballistic regime. It is a remarkable fact that smaller values do not give a big difference with the clean case, for times up to $t=1000$. 

\begin{figure}
  \centering
  \includegraphics[width=0.9\linewidth]{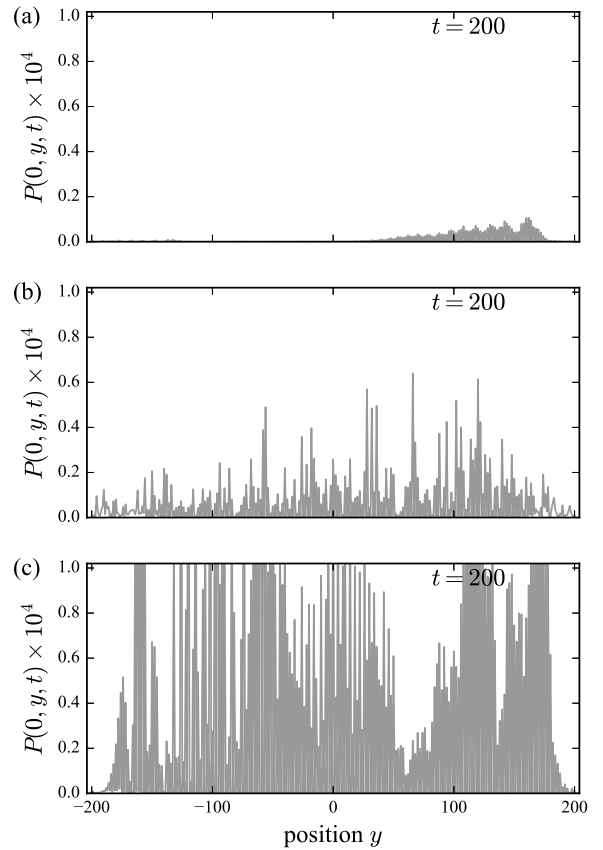}
  \caption{Short time ($t=200$) position distribution of a quantum walk for an initial condition at $x=-10$, in the spin up state. (a) Clean case, the walk freely propagates on the lattice; (b) with rotation disorder it is localized ($J=0.2$, $p=0.03$),  and (c) with nonlinear disorder it is partially trapped at the interface restoring the ballistic propagation ($J=100$, $p=0.1$). Only the nonlinear quantum walk is trapped at the interface.
  \label{f:chern_x0}}
\end{figure}

\begin{figure*}
  \centering
  \includegraphics[width=0.5\linewidth]{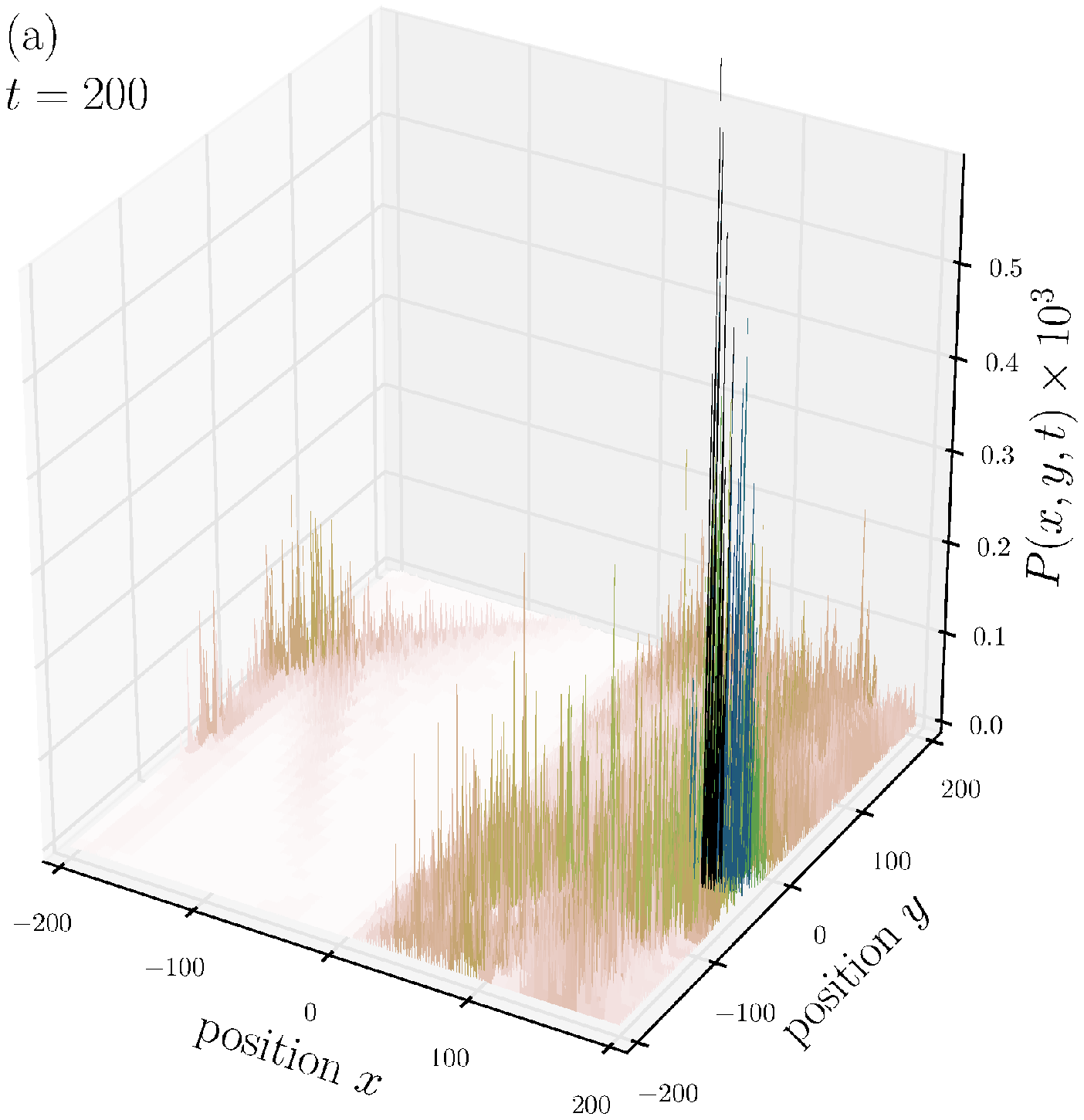}%
  \includegraphics[width=0.5\linewidth]{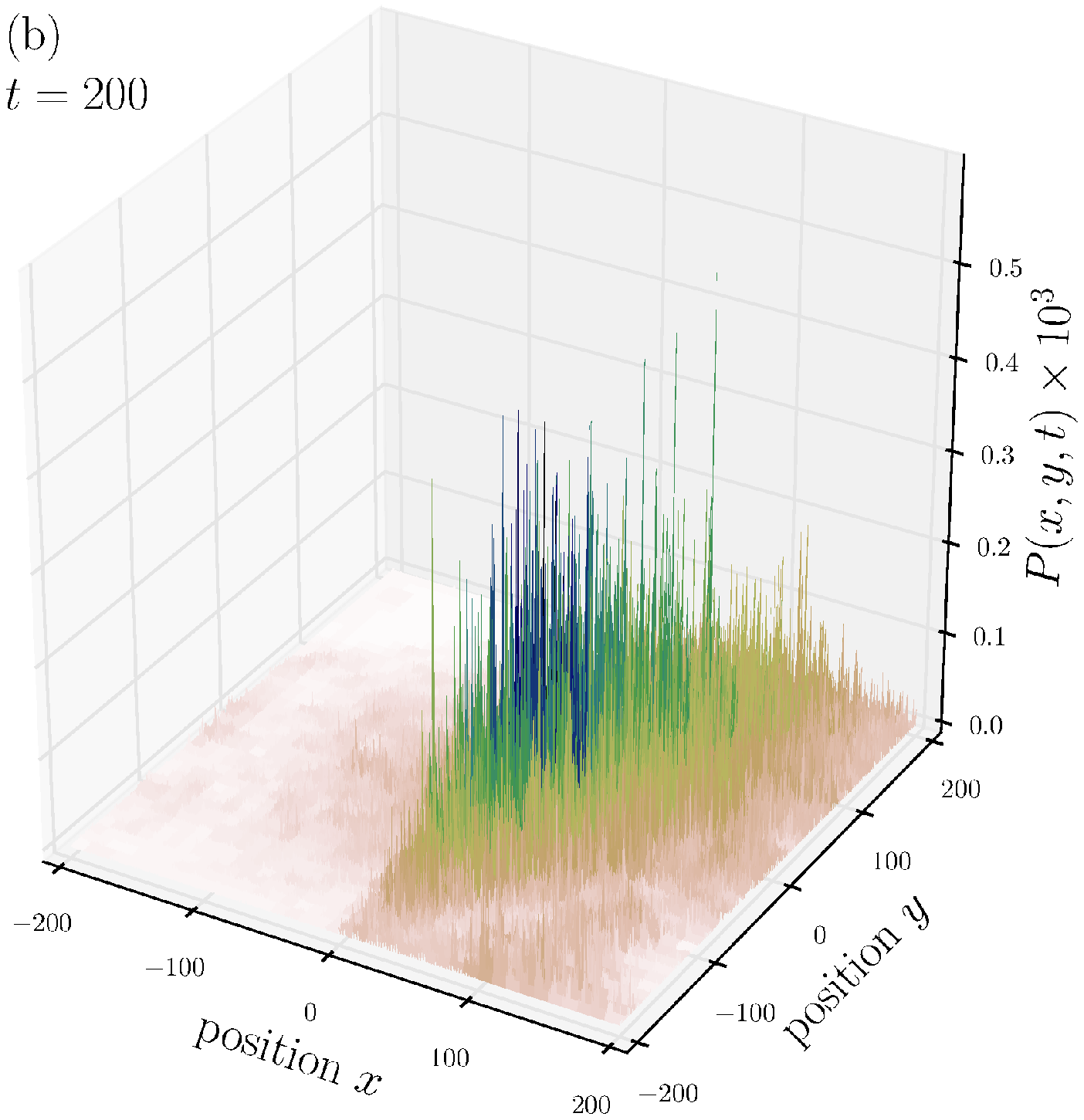}\\
  \includegraphics[width=0.5\linewidth]{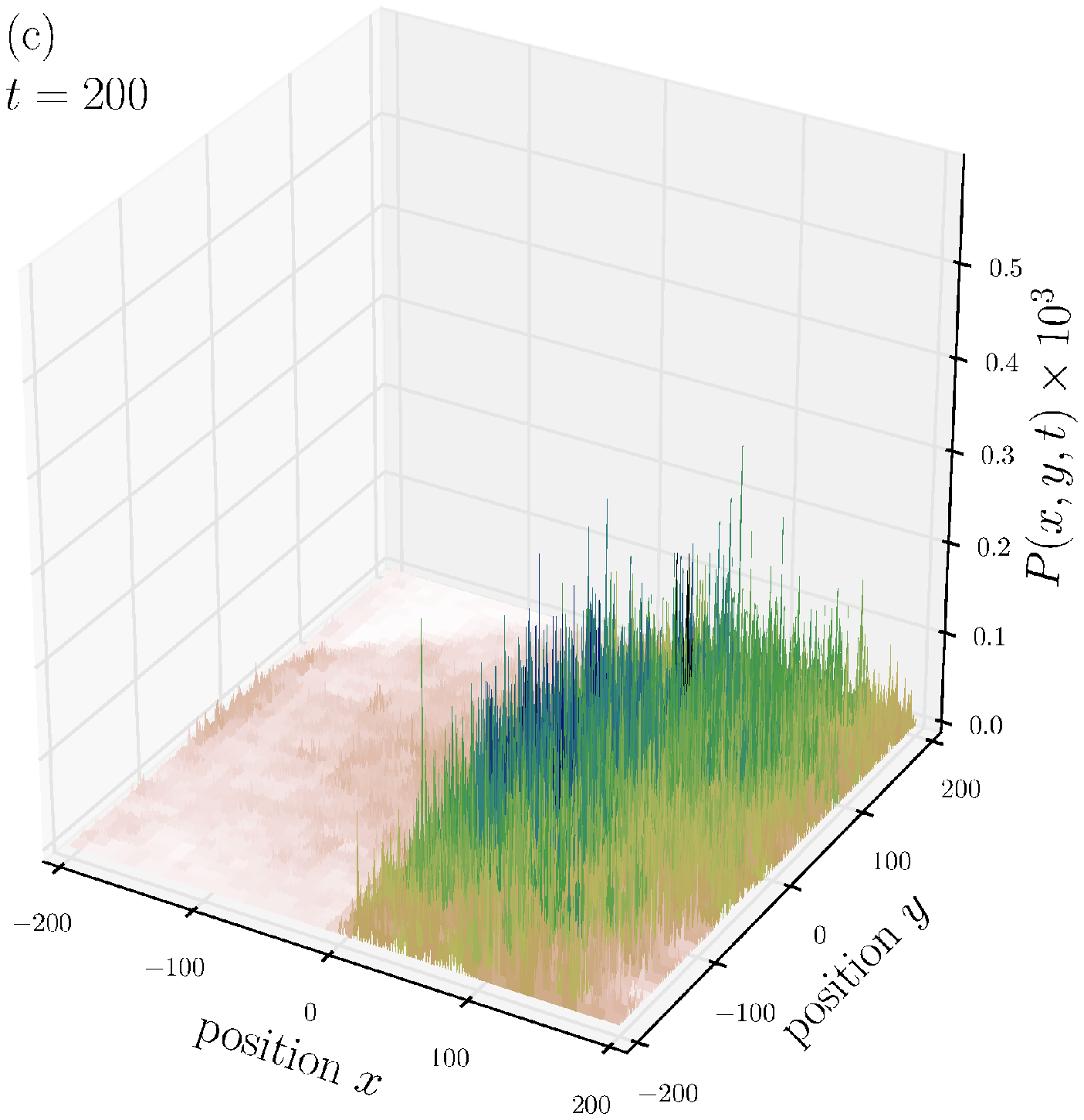}%
  \includegraphics[width=0.5\linewidth]{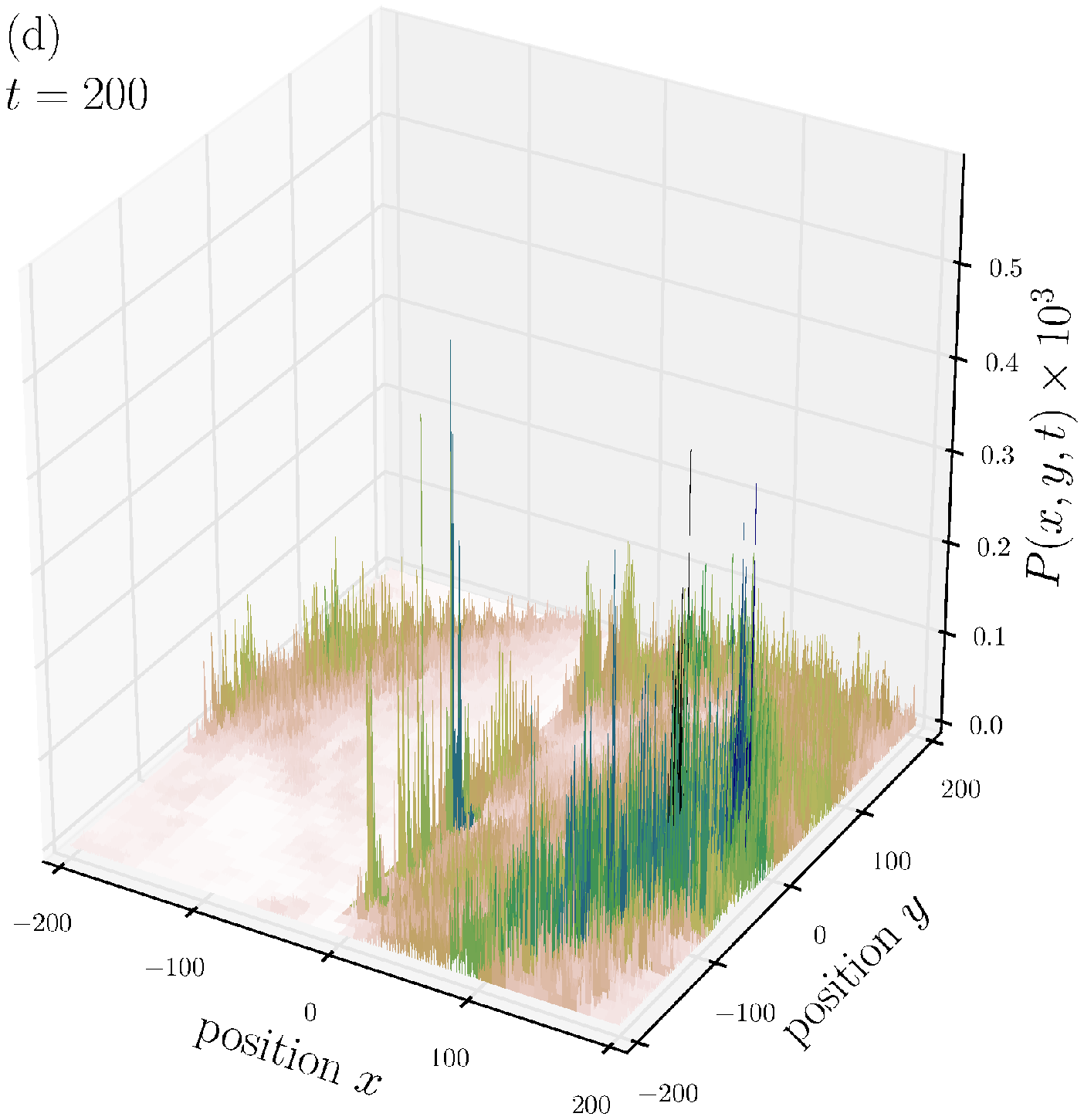}
  \caption{Position distribution for a walk started at $x=-10$, outside the interface, in a spin up state (thus, mainly propagating towards the right); (a) clean, (b) rotation disorder ($p=0.03$, $J=0.2$), (c) phase disorder ($p=0.03$, $J=0.2$), and (c) nonlinear disorder ($p=0.1$, $J=100$).
   \label{f:chern_x_01}}
\end{figure*}

The most striking observation is that the ballistic transport is fully preserved, even in the direction perpendicular to the interface. As shown in Fig.~\ref{f:chern_0809}, where we plot the density and walk width at the interface in the case of nonlinear disorder, we observe ballistic spreading proportional to $t$; though not shown in the figure, we find in particular that the isotropic width, after an initial transient, is also linear in time.

Merely for extreme values of the coupling constant $J \gg 1/p$, the well structured interference pattern of the clean walk tends to loss its organization. This start to be visible on the small scale amplitudes as shown on the three panels of Fig.~\ref{f:chern_0809s}, where we compare the spin density for the clean case with the nonlinear walk for two values of $J$. For smaller values of $J$ the nonlinear walk behavior remains close to the clean one, for the same initial condition starting at the interface. In sharp contrast with the ballistic transport of the nonlinear walk, if we replace the self-consistent phases \(\phi=\phi(\psi)\) in \eqref{e:phi} by random angles \(\phi\sim \mathcal{U}(0,2\pi)\), we numerically observe that the edge channel is broken and the walk cannot spread at a ballistic rate (Fig.~\ref{f:chern_0809}c). This can be understood as a consequence of the random phase shifts in the $z$ direction that for strong enough disorder lead to spin flips hence changing locally the walker's direction. By the way of comparison we remark the special behavior of the system governed by the nonlinear coin in which the edge states appear to be robust. Therefore, the nonlinear disorder restores the ballistic behavior lost for random phases, even for much larger values of the coupling constant.

In fact, the edge state not only provides a channel to transport the information, it is also an attractor to the nonlinear walk. Indeed, if in the absence of disorder one starts with an initial state that does not overlap with the edge state, for instance at $x=-10$, the quantum walk propagates freely on the whole lattice. However, adding the nonlinear interaction, the walk is partially trapped at the interface where it propagates at a ballistic rate, in spite of the spatial disorder (Figs.~\ref{f:chern_x0} and \ref{f:chern_x_01}). In Fig.~\ref{f:chern_x0} we compare at $t=200$ the density distribution at the interface, and in Fig.~\ref{f:chern_x_01} the distribution over the lattice. The initial particle position is chosen in the left region ($x=-10$ at $t=0$), and the spin set up, to favor the propagation towards the right, allowing the walker encounter the interface. Figure~\ref{f:chern_x_01} clearly shows the difference between the clean case (a), the random rotation angles (b) and phases (c) disordered cases, and finally the nonlinear walk (d). The anisotropy is a consequence of the choice of the initial state (spin up). Only in the nonlinear case an accumulation of probability density on the interface is observed. This behavior, for which the interface edge states act as an attractor, is reminiscent to a generalization to two dimensions, of the observed phenomenon of localization at a topological defect in one dimension \cite{Gerasimenko-2016fk}. This is related to the fact that at low energy and for small values of the rotation angle, the one dimensional walk limits to the nonlinear Dirac equation that has soliton solutions \cite{Lee-2015uq}.

Indeed, using the explicit expression of the evolution operator in momentum space, it is easy to obtain its ``hydrodynamic'' limit \(\bm k \rightarrow 0\), \(U\approx u_0 + u_x \I k_x + u_y \I k_y\):
\begin{align}
  u_0 & = R\left(\theta+ \frac{\alpha}{2} \right)\,, \nonumber \\
  u_x & = 2 \cos \frac{\theta+\alpha}{2} \left( -\sin \frac{\theta}{2} \sigma_x +
  \cos \frac{\theta}{2} \sigma_z \right) \nonumber \\
  u_y & = 2 \cos \frac{\theta}{2} \sigma_z
  \label{e:uoxy}
\end{align}
which leads, using a crude approximation for small angles
$$
u_0 \approx \sigma_0 - \I \left( \theta + \frac{\alpha}{2} \right) \sigma_y,\;
u_x \approx 2\sigma_z - \frac{\theta}{2} \sigma_x, \;
u_z \approx 2 \sigma_z\,,
$$
and adding the nonlinear term in \(J\) as a smooth mean-field,
$$
\overline{s}(\bm x, t) = \langle \psi^\dagger| \sigma_z | \psi \rangle\,,
$$
(averaged spin distribution) to the Dirac like equation in \(2+1\) dimensions:
\begin{multline}
  \label{e:dirac}
  \left[ 
    \frac{\partial}{\partial t} - 2 \sigma_z \left(
      \frac{\partial}{\partial x} + \frac{\partial}{\partial y}
    \right) + \theta \sigma_x \frac{\partial}{\partial x}
  \right] \psi(\bm x, t) + \\
  \I \left[ \left(
    \theta + \frac{\alpha}{2} \right) \sigma_y + 
    g \overline{s}(\bm x, t) \sigma_z
  \right] \psi(\bm x,t) = 0\,,
\end{multline}
where \(\psi\) is the position representation of the spinor, and $g = 3\pi p J$ the effective coupling constant (cf. \cite{Lee-2015uq,Molfetta-2012fv} for a more formal calculation in the one dimensional case). In addition to the ``mass'' term which gives the boundary separating two different topologies (the mass vanishes for \(\alpha + 2\theta = 0\), in the present approximation for small angles). The effective coupling constant is of the order of $pJ$, which justifies the choice of $J\sim1/p$ as the order of magnitude of the nonlinear noise in the numerical computations. The term in $\theta \partial_x$ is the lowest order term showing the anisotropy of the walk, the ``velocity'' $\theta$ being smaller than the ``light velocity'' $c=2$ in the diagonal direction. We note that the nonlinearity appears as a self-consistent gauge field. The nonlinear term, being proportional to $\I\sigma_z$, adds naturally to the first order spatial derivatives also proportional to $\sigma_z$, in a way similar to a ``vector'' potential \cite{Jackiw-1984fk}. This is somewhat different to the more usual nonlinear Gross-Neveu model in which the nonlinear term adds to the mass term \cite{Lee-2015uq}, or the state dependent rotation of Ref.~\cite{Gerasimenko-2016fk} which preserves the particle-hole symmetry. 

%
\section{Conclusion}
In conclusion, we investigated the effect of spatial disorder in a two dimensional discrete quantum walk. We demonstrated the ballistic propagation along the edge state localized at the interface between two distinct topological phases. This edge state is robust against quenched disorder, provided it cannot change the system's topology. In particular, we found an anisotropic state with a bulk localized density distribution coexisting with ballistic propagation at the interface. 

In the case of a coin operator depending on nonlinear phases randomly distributed in space, the ballistic propagation of the information, mostly as in the absence of disorder, is preserved even for very strong couplings. This is also in contrast with the behavior of a walk perturbed by random phases breaking the particle-hole symmetry, in which case the ballistic propagation is no longer possible. In addition a new phenomenon arises, the walk can be trapped at the interface. The nonlinear walk can be mapped to a Dirac equation in the continuous limit with a mass term depending on the rotation angles and a nonlinear term proportional to the smoothed spin density, known to possess propagating localized solutions along the interface (the transverse direction being localized).

We focused here on the rich and striking phenomenology of quantum walks and their topological edge states; the analysis of the mechanisms behind the observed effects of disorder, especially in the nonlinear case, deserves further consideration.

\begin{acknowledgement}
We benefited from useful discussions with Laurent Raymond and Thomas Krajewski. We thank Giuseppe di Molfetta for his interest and thorough comments. This work was partially supported by CNRS UMR 7332, and Université de Toulon.
\end{acknowledgement}

\bibliographystyle{epj}

\end{document}